\documentclass[a4paper,11pt]{article}

\usepackage[english]{babel}

\usepackage[a4paper,tmargin=3truecm,bmargin=3truecm,rmargin=3truecm,
lmargin=3truecm,twoside,verbose=true]{geometry}

\usepackage{cancel,graphicx}
\usepackage[pdftex]{hyperref}
\usepackage{amsmath,amssymb}

\numberwithin{equation}{section}

\allowdisplaybreaks[1]

\usepackage{cancel,graphicx}
\usepackage{hyperref}
\usepackage{amsmath,amssymb,slashed}

\usepackage{amsmath,amssymb}
\usepackage[amsmath, hyperref, thmmarks]{ntheorem}
\numberwithin{equation}{section}

\allowdisplaybreaks[1]

\newcommand\bbbone{\mathbb{I}}

\newcommand\del{{\partial}}

\newcommand{\eqn}[1]{(\ref{#1})}
\newcommand{\R}{\mathbb{R}}

\newcommand{\be}{\begin{equation}}
\newcommand{\ee}{\end{equation}}

\theoremsymbol{}
\theorembodyfont{\slshape}
\theoremheaderfont{\normalfont\bfseries}
\theoremseparator{}

\theorembodyfont{\upshape}
\theoremsymbol{\ensuremath{\blacklozenge}}

\theoremstyle{nonumberplain}
\theoremheaderfont{\scshape}
\theorembodyfont{\normalfont}
\theoremsymbol{\ensuremath{\blacksquare}}

\qedsymbol{\ensuremath{_\blacksquare}}
\theoremclass{LaTeX}

\renewenvironment{thebibliography}[1]
         {\section*{References}\frenchspacing\small
          \begin{list}{[\arabic{enumi}]}
         {\usecounter{enumi}\parsep=2pt\topsep 0pt
         \settowidth{\labelwidth}{[#1]}
         \leftmargin=\labelwidth\advance\leftmargin\labelsep
         \rightmargin=0pt\itemsep=1pt\sloppy}}{\end{list}}
\title{Noncommutative gauge theories on $\mathbb{R}^2_\theta$ as matrix models}
\author{Pierre Martinetti$^{a,b}$, Patrizia Vitale $^{a,b}$ and Jean-Christophe Wallet$^c$}

\begin{document}
\date{}
\maketitle
\vspace*{-1cm}

\begin{center}
\textit{$^a$Dipartimento di Fisica, Universit\`a di Napoli Federico II\\
$^b$ INFN Sezione di Napoli, Via Cintia 80126 Napoli, Italy\\
    e-mail: 
\texttt{martinetti.pierre@gmail.com, vitale@na.infn.it}}\\[1ex]
\textit{$^c$Laboratoire de Physique Th\'eorique, B\^at.\ 210\\
CNRS and Universit\'e Paris-Sud 11,  91405 Orsay Cedex, France\\
    e-mail: 
\texttt{jean-christophe.wallet@th.u-psud.fr}}\\[1ex]

\end{center}

\begin{abstract}
We study a class of noncommutative gauge theory models on 2-dimensional Moyal space from the viewpoint of matrix models and explore some related properties. Expanding the action around symmetric vacua generates non local matrix models with polynomial interaction terms. For a particular vacuum, we can invert the kinetic operator which is related to a Jacobi operator. The resulting propagator can be expressed in terms of Chebyschev polynomials of second kind. We show that non vanishing correlations exist at large separations. General considerations on the kinetic operators stemming from the other class of symmetric vacua, indicate that only one class of symmetric vacua should lead to fast decaying propagators. The quantum stability of the vacuum is briefly discussed.
\end{abstract}
\vskip 1 true cm
\noindent Keywords: {\it{Noncommutative Geometry; noncommutative gauge theories; matrix models}.}
\pagebreak

\section{Introduction.}

In Noncommutative Geometry (NCG) \cite{Connes1}, one of the  guiding  ideas is to generalize the  duality, existing in Riemannian geometry,    between spaces and associative algebras in a way that the structural properties of the space, for instance topological, metric, differential properties, can be given an algebraic description. It turns out that many of the building blocks of modern physics fit well with the basic concepts of NCG which may ultimately lead to a better understanding of spacetime at short distance. For instance, NCG provides a  way to resolve  the physical objections, emerging from the concurrency of General Relativity and Quantum Mechanics  \cite{Doplich1},  to the existence of  continuous space-time  at  Planck scale. Once the noncommutative nature of space(-time) is assumed, it is natural to consider field theories on noncommutative spaces, called Noncommutative Field Theories (NCFT). NCFT emerged around 1986 in String field theory \cite{witt1}, followed a few years later  by the first NCFT's on the fuzzy sphere, a finite dimensional NCG, \cite{fuzzy1}. NCFT on the Moyal space were singled out as effective regimes of String theory \cite{Schomerus, sw99} around 1998 and as underlying structures in quantum Hall physics \cite{qhe1}, \cite{qhe4}. For reviews on Moyal NCFT, see e.g  \cite{dnsw-rev}.\par

The renormalization of NCFT is difficult, unless the underlying NCG is finite. In Moyal geometry, this is due to the ultraviolet/infrared (UV/IR)  mixing, appearing already in the real-valued $\varphi^4$ model on the 4-d Moyal space \cite{Minwalla:1999px}. It comes from UV finite nonplanar diagrams which exhibit IR singularities generating new divergences when inserted into higher order diagrams that cannot be cured. A first solution to the mixing in the scalar field theory amounts to add to the initial action a harmonic oscillator term. This yields the Grosse-Wulkenhaar model, perturbatively renormalizable to all orders \cite{Grosse:2003aj-pc}, \cite{Grosse:2003-r2}, \cite{grWm}. Various related properties have been studied, among which classical and/or geometrical ones, $2$-d fermionic extensions \cite{hghs}-\cite{LVW-07}. The Grosse-Wulkenhaar model has vanishing of the $\beta$-function to all orders \cite{beta11} when it is self-dual under the Langmann-Szabo duality \cite{LSD}, and is very likely to be non-perturbatively solvable \cite{harald-raimar}. Scalar field theories on the noncommutative space $\mathbb{R}^3_\lambda$, a deformation of $\mathbb{R}^3$, which are free of UV/IR mixing have been built recently in \cite{vit-wal-12}. Whether this simply comes from the low ``dimension'' of the space or reflects a specific property of the underlying NCG remains to be seen{\footnote{A  discussion on the  origin of the mixing for translation-invariant products can be found in \cite{gallu-1}. }} .\par

The UV/IR mixing also occurs in gauge models on 4-d Moyal space \cite{Matusis:2000jf}. For early studies, see e.g \cite{gauge-var113} and references therein. The mixing appears in the naive noncommutative version of the Yang-Mills action given by $S_0=\frac 14\int d^4x (F_{\mu\nu}\star F_{\mu\nu})(x)$, showing up at one-loop order as a hard IR transverse singularity in the vacuum polarization tensor. Attempts to extend the Grosse-Wulkenhaar harmonic solution to a gauge theoretic framework have singled out \cite{GWW} a gauge invariant action expressed as{\footnote{Notations and conventions are collected in the section \ref{section2}.}} 
\begin{align}
S_{\Omega}=\int d^4x \Big(\frac 14F_{\mu\nu}\star F_{\mu\nu}
+\frac{\Omega^2}{4}\{\mathcal A_\mu,\mathcal A_\nu\}^2_\star
+\kappa\mathcal A_\mu\star\mathcal A_\mu\Big) \label{inducedgauge}.
\end{align}
where $\Omega$ and $\kappa$ are real parameters, while  $\mathcal A_\mu=A_\mu+\frac{1}{2}\tilde x_\mu$ is a gauge covariant one-form given by the difference of the gauge connection and the natural   gauge invariant connection  induced by the minimal  (e.g. based on the derivations $\del_\mu$)  Moyal differential calculus \cite{Dubois} (see section \ref{section2}). The corresponding mathematical framework has been developed in \cite{WAL1, WAL3}. Eq. \eqref{inducedgauge} can also be viewed as a spectral action related to a so called finite volume spectral triple \cite{harmonic-gw}, whose Dirac operator is a square root of the kinetic operator of the 4-dimensional Grosse-Wulkenhaar model. The related noncommutative (metric) geometry is rigidly linked to the Moyal (metric) geometries, as shown in \cite{homot-moyal, moyal1,moyal2}. Unfortunately, the action \eqref{inducedgauge} is hard to deal with when it is viewed as a functional of the gauge potential $A_\mu$, $S_\Omega[A_\mu]$. This is mainly due to its complicated vacuum structure explored in \cite{GWW2} which excludes the use of any standard perturbative treatment. Other attempts to control the UV/IR mixing have been considered in \cite{Blaschke:2009aw} -\cite{Blaschke:2009c}. However,  showing that any of these models is renormalizable is still an open problem.\par

When expressed as a functional of the covariant one-form ${\cal{A}}_\mu$,  the action  \eqref{inducedgauge} bears some similarity with a matrix model, where the field  ${\cal{A}}_\mu$ can be  represented as an infinite matrix in  the Moyal matrix base (see Appendix \ref{matrixbase}). 
To our knowledge this interpretation has not been explored so far for the action \eqn{inducedgauge}, although the matrix model formulation of NC gauge theory has been known since many years \cite{matrix}. 
Keeping in mind that ${\cal{A}}_\mu$ is a natural variable of the Moyal geometry \cite{WAL1}, a sensible question is to explore the properties of the action as a functional of the field   ${\cal{A}}_\mu$, in order to determine to what extent $S_\Omega[{\cal{A}}_\mu]$ may give rise to a meaningful quantum theory.

This requires at least to choose a vacuum, expand the action around it, and perform the difficult but mandatory computation of the propagator. In the 4-d Moyal case, an additional complication comes from the need to control the UV/IR mixing behavior of the ghost sector coupled to the gauge sector. This can be easily overcome in the 2-d Moyal case, that we will consider in this paper, thanks to a suitable gauge choice, akin to the temporal gauge, for which the ghosts decouple from the gauge sector.\par

The purpose of this paper is to perform a first exploration of the gauge theory  on 2-dimensional Moyal space described by the action $S_\Omega[{\cal{A}}_\mu]$  of Eq. \eqref{inducedgauge},  at $\Omega\ne 0$,  viewed as a matrix model as explained above. The expansion of the action around the non-trivial symmetric vacua determined in \cite{GWW2}, to which we restrict in this paper, and a BRST further gauge-fixing give rise generically to a non local matrix model with a complicated kinetic operator together with cubic and quartic polynomial interaction terms. For a particular symmetric vacuum corresponding to $\Omega=\frac{1}{3}$, we show that the kinetic operator is a Jacobi operator. The computation of the propagator can then be carried out. The resulting propagator can be expressed in terms of Chebyschev polynomials of second kind. We  show that non vanishing correlations exist at large separations.  The quantum stability of the vacuum is briefly discussed in the particular situation. A divergent 1-point function appears signaling likely a loss of symmetry at the quantum level.

Extending our analysis to general considerations on the kinetic operators stemming from the other   vacua found in \cite{GWW2}, we single out a particular class of symmetric vacua that should lead to fast decaying propagators at large separation, reminiscent of a quasi-local behaviour for the corresponding matrix models.\par

In  section \ref{section2} we collect relevant properties of Moyal geometry that we will need together with specific features of  the action \eqref{inducedgauge} at $\Omega=0$, such as the absence of UV/IR mixing and additional cancellation between IR singularities specific to 2 dimensions. The case $\Omega\ne0$ is considered in  section \ref{section3} which involves the general expansion and gauge fixing together with the computation of the propagator. We finally discuss the results and conclude.\par

\section{\texorpdfstring{Yang-Mills type theory on $\mathbb{R}^2_\theta$: The $\Omega=0$ case.}{The Yang-Mills case}}\label{section2}
\subsection{Basics on Moyal algebra and gauge invariant actions.}\label{use}
In this subsection  we collect useful properties of the Moyal algebra. For more details, see e.g \cite{Gracia-Bondia:1987kw}. Let ${\cal{S}}({\mathbb{R}}^2)\equiv{\cal{S}}$ and ${\cal{S}}^\prime({\mathbb{R}}^2)\equiv{\cal{S}}^\prime$ be respectively 
the space of complex-valued Schwartz functions and the dual space of tempered distributions on ${\mathbb{R}}^2$. The  associative Moyal $\star$-product is defined for all $f, g$ in ${\cal{S}}$ by the map: $\star:{\cal{S}}\times{\cal{S}}\to{\cal{S}}$ 
\begin{equation}
(f\star g)(x)=\frac{1}{(\pi\theta)^2}\int d^2y\,d^2z\ f(x+y)g(x+z)e^{-i\,2y^\mu\,\Theta^{-1}_{\mu\nu}z^\nu},\ \Theta_{\mu\nu}=\theta\begin{pmatrix} 0&1 \\ -1& 0 \end{pmatrix} \label{moyal}
\end{equation}
where $\theta\in\mathbb{R}$, $\theta>0${\footnote{ We will use the notation $y\Theta^{-1}z\equiv y_\mu \Theta^{-1}_{\mu\nu}z^\nu$ and Einstein summation convention.}}. The integral is a faithful trace: $\int d^2x\ (f\star g)(x)=\int d^2x\ (g\star f)(x)=\int d^2x\ f(x)g(x)$. The Leibniz rule holds: $\partial_\mu(f\star g)=\partial_\mu f\star g+f\star\partial_\mu g$, $\forall f,g\in{\cal{S}}$. The complex conjugation in ${\cal{S}}$, $a\mapsto a^\dag$, $\forall a\in{\cal{S}}$, defines an involution in ${\cal{S}}$. It extends to ${\cal{S}}^\prime$ by using duality of vector spaces. The $\star$-product \eqref{moyal} can be extended to ${\cal{S}}^\prime\times{\cal{S}}\to{\cal{S}}^\prime$ using again duality
($\langle T\star a,b \rangle = \langle T,a\star
b\rangle$, $\forall T\in{\cal{S}}^\prime$, $\forall a,b\in{\cal{S}}$) and continuity of \eqref{moyal}. In a similar way, \eqref{moyal} extends
to ${\cal{S}} \times {\cal{S}}^\prime\to{\cal{S}}^\prime$, via $\langle a\star T,b \rangle = \langle T,b\star
a\rangle$, $\forall T\in{\cal{S}}^\prime$, $\forall a,b\in{\cal{S}}$.\\
Here, the Moyal algebra denoted by $\mathbb{R}^2_\theta$ is the multiplier algebra of $({\cal{S}},\star)$,  $\mathbb{R}^2_\theta:={\cal{M}}_L\cap{\cal{M}}_R\label{Moyal-alg}$
where  ${\cal{M}}_L=\{T\in{\cal{S}}^\prime\ /\ a\star T\in{\cal{S}},\ \forall a\in{\cal{S}}\}$ and ${\cal{M}}_R=\{T\in{\cal{S}}^\prime\ /\ T\star a\in{\cal{S}},\ \forall a\in{\cal{S}}\}$ are respectively the left and right multiplier algebras of $({\cal{S}},\star)$ \cite{Gracia-Bondia:1987kw}. We set $[a,b]_\star:= a\star b-b\star a$. For any $a,b\in\R^2_\theta$, the following relations hold true:
\begin{eqnarray}
\partial_\mu(a\star b)&=&\partial_\mu a\star b+a\star\partial_\mu
b,\qquad \;\; \;\;
(a\star b)^\dag=b^\dag\star a^\dag,\qquad \label{relations1}
\\
x_\mu\star a&=&(x_\mu\cdot a)+{{i}\over{2}}\Theta_{\mu\nu}\partial_\nu a,\qquad a\star\ x_\mu=(x_\mu\cdot a)-{{i}\over{2}}\Theta_{\mu\nu}\partial_\nu a\label{relations2}.
\end{eqnarray}
Note that Eq. \eqref{relations2} implies the celebrated relation $[x_\mu,x_\nu]_\star =i\Theta_{\mu\nu}$ among the coordinate functions of $\mathbb{R}^2_\theta$. In  section \ref{section3}, we will use the matrix base of $\mathbb{R}^2_\theta$ \cite{Gracia-Bondia:1987kw}. The relevant material is given in  appendix \ref{matrixbase}.\par

As shown in \cite{WAL1}, the algebraic properties of classical gauge invariant actions on the Moyal plane are described by a simple version of the derivation-based differential calculus. Here, the unital algebra itself , $\mathbb{R}^2_\theta$,  plays the role of the right $\mathbb{R}^2_\theta$-module while the Lie algebra of derivations is the Abelian algebra generated by the  derivatives $\partial_\mu$, $\mu=1,2$. The map $\nabla_X:\mathbb{R}^2_\theta\to\mathbb{R}^2_\theta$, $X=(\partial_\mu),\ \mu=1,2$ given  by
\begin{equation}
\nabla_{\partial_\mu}(a)=\nabla_\mu(a)=\partial_\mu a-iA_\mu\star a,\ ;\;\; A_\mu=i\nabla_\mu(\bbbone_{\mathbb{R}^2_\theta}),\;\;\; \forall a\in\R^2_\theta\label{nablamiu}
\end{equation}
with $A_\mu^*=A_\mu$ defines a hermitian connection, $\nabla$,  for the hermitian structure  
$h:\mathbb{R}^2_\theta\times\mathbb{R}^2_\theta\to\mathbb{R}^2_\theta,\ h(m_1,m_2)=m_1^*\star m_2, \forall m_1,m_2\in\mathbb{R}^2_\theta$.
The curvature of $\nabla$ is
\begin{equation}
R(X,Y):\mathbb{R}^2_\theta\to\mathbb{R}^2_\theta,\ R(X,Y)m=([\nabla_X,\nabla_Y]-\nabla_{[X,Y]})m,\ \forall m\in\mathbb{R}^2_\theta\label{generalcurvature}
\end{equation}
for any derivations $X$ and $Y$ and in the present case yields
\begin{equation}
R_{\mu\nu}:=R(\partial_\mu,\partial_\nu)=-i(\partial_\mu A_\nu-\partial_\nu A_\mu-i[A_\mu,A_\nu]_\star)\label{curvature}.
\end{equation}
To make contact with usual  notation, we set from now on $R_{\mu\nu}=-iF_{\mu\nu}$.
The unitary gauge group ${\cal{U}}(\mathbb{R}^2_\theta)$ acts  as $\nabla_X^g=g^\dag\circ\nabla_X\circ g$. This yields
\begin{equation}
A_\mu^g=g^\dag\star A_\mu\star g+ig^\dag\star\partial_\mu g,\ F_{\mu\nu}^g=g^\dag\star F_{\mu\nu}\star g,\;\;\;\;\; \forall g\in{\cal{U}}(\mathbb{R}^2_\theta)\label{gaugetrans}.
\end{equation}
We define ${\tilde{x}}_\mu:=2\Theta^{-1}_{\mu\nu}x_\nu$. Recall \cite{WAL1} that  the one-form  components $A_\mu=-{{1}\over{2}}{\tilde{x}}_\mu:=A_\mu^{inv}$, define a gauge invariant connection one-form so that the map \eqref{nablamiu} becomes:
\begin{equation}
\nabla^{inv}_\mu(a)=\partial_\mu a+{{i}\over{2}}{\tilde{x}}_\mu\star a={{i}\over{2}}a\star{\tilde{x}}_\mu,\ \forall a\in\mathbb{R}^2_\theta\label{gaugeinvconnection},
\end{equation}
where gauge invariance $(\nabla^{inv}_X)^g=g^\dag\circ\nabla^{inv}_X\circ g=\nabla^{inv}_X$ follows from the second equality in Eq.  \eqref{gaugeinvconnection}. The corresponding curvature is $F^{inv}_{\mu\nu}=-i\Theta^{-1}_{\mu\nu}$. Thanks to the existence of the gauge invariant connection, a natural covariant one-form can be defined as 
\begin{equation}
{\cal{A}}_\mu:=i(\nabla_\mu-\nabla^{inv}_\mu)=A_\mu+{{1}\over{2}}{\tilde{x}}_\mu,\; \;\;  {\cal{A}}_\mu^g=g^\dag\star{\cal{A}}_\mu\star g,\;\;\; \forall g\in{\cal{U}}(\mathbb{R}^2_\theta)\label{covcoordinate}.
\end{equation}
${\cal{A}}_\mu$ are formally  the ``covariant coordinates'' introduced in  \cite{MSSW2000} on Moyal space-time in the presence  of a local symmetry, although here they play a different role, namely that of the dynamical gauge variables. From Eq. \eqref{covcoordinate} and the first of Eqs.  \eqref{relations2}, one obtains
\begin{equation}
F_{\mu\nu}=-i[{\cal{A}}_\mu,{\cal{A}}_\nu]_\star+\Theta^{-1}_{\mu\nu}\label{bracketform}.
\end{equation}
The extension to $\mathbb{R}^{2n}_\theta$, $n\in\mathbb{N}$ is straightforward. For $n=2$, the above framework underlies the action \eqref{inducedgauge} which can be expressed solely in terms of ${\cal{A}}_\mu$ \cite{GWW}, namely:
\begin{equation}
S_{\Omega}=\int d^4x\ \Big(-\frac{1}{4}[{\cal{A}}_\mu,{\cal{A}}_\nu]^2_\star+\frac{\Omega^2}{4}\{\mathcal A_\mu,\mathcal A_\nu\}^2_\star
+\kappa\mathcal A_\mu\star\mathcal A_\mu\Big) \label{inducedgaugematrix}.
\end{equation}
The gauge invariance is obvious in view of the tracial property of the integral and the 2nd relation in \eqref{covcoordinate}. Note that this action can be obtained from a spectral triple of a specific type introduced in \cite{harmonic-gw} and shown, in the sense of noncommutative metric spaces,  to be homothetic \cite{homot-moyal} to the standard Moyal spectral triple. \par 

\subsection{\texorpdfstring{The $\Omega=0$ case.}{The Yang-Mills case}}\label{subsection2.2}

When the real parameters $\Omega$ and $\kappa$ are set to zero, the action \eqref{inducedgaugematrix}, expressed as a functional of $A_\mu$, exhibits  UV/IR mixing which shows up at the one-loop level as a hard IR singularity in the vacuum polarization tensor. The situation is more favorable for the 2-d version of \eqref{inducedgaugematrix}. The corresponding model on $\mathbb{R}^2_\theta$ at $\Omega=\kappa=0$ is known to be UV/IR mixing free. Only the sector of planar diagrams actually matters and this latter is similar to the perturbative expansion of a commutative 2-d Yang-Mills theory. In fact, by using a ``temporal-like'' gauge, e.g  $A_2=0$, the gauge fixed action is purely quadratic and separates into a gauge potential $A_\mu$ part and a ghost part, as for commutative 2-d Yang-Mills theories. Alternatively, one may use the popular ``covariant'' Landau gauge for which the gauge fixed theory ``does not look free''. By standard calculation, it can be easily realized that the hard as well as logarithmic IR singularities in the vacuum polarization tensor responsible for the UV/IR mixing are proportional to $(2-d)$ \cite{WAL1}, \cite{Blaschke:2008phd}. This cancellation propagates to other higher order correlation functions as a consequence of Slavnov-Taylor identities \cite{hgw-13}. We close this section by recalling that massless 2-d theories are known to exhibit additional IR singularities. In the case of 2-d commutative Yang-Mills theories, these depend of the gauge function. It is instructive to examine more closely the fate of these additional singularities within $S_{\Omega=0}[A_\mu ]$ by using the Landau gauge. The gauge fixed action is $S_{tot}=S_{\Omega=0}[A_\mu ]+S_{GF}$ with
\begin{equation}
S_{GF}=s\int d^2x\big({\bar{C}}\partial^\mu A_\mu\big)=\int d^2x\big(b\partial^\mu A_\mu-{\bar{C}}\partial^\mu(\partial_\mu C-i[A_\mu,C]_\star)\big)\label{sgflandau}
\end{equation}
where the Slavnov operation $s$ is defined by $sA_\mu=\partial_\mu C-i[A_\mu,C]_\star,\quad sC=iC\star C,\quad s{\bar{C}}=b,\quad sb=0$. Here, $C$, ${\bar{C}}$ and $b$ are respectively the ghost, the antighost and the St\"uckelberg field with ghost number equal to $+1$, $-1$ and $0$. $s$ acts as a graded derivation with grading defined by the sum of the degree of forms and ghost number (modulo 2) and $s^2=0$. \par 

Standard calculation using \eqref{vcc} and \eqref{JN} leads to the planar ($\omega_g^{P}(p)$) and non planar ($\omega_g^{NP}(p)$) 1-loop contributions to the ghost 2-point function:
\begin{equation}
\omega_g^{P}(p)={2^{1-{{D}\over{2}} }\over{(2\pi)^{D/2}(p^2)^{{{D}\over{2}}-1} }} 
({{\Gamma({{D}\over{2}})\Gamma({{D}\over{2}}-1) }\over{\Gamma(D-1) }})\Gamma(2-{{D}\over{2}})\label{omegaghostplanaire2},
\end{equation}
\begin{equation}
\omega_g^{NP}(p)=-2p^2\int_0^1dxx{{1}\over{ 
(2\pi)^{D/2}}}(M^2)^{{{D}\over{2}}-2}\big( {{1}\over{2}} ({\sqrt{ {\tilde{p}}^2M^2}})^{2-{{D}\over{2}}}
K_{2-{{D}\over{2}}}({\sqrt{ {\tilde{p}}^2M^2}})\big)\label{omegaghostnonplan},
\end{equation}
where $M^2:=p^2x(1-x)$.
For $D=2$, the UV finite planar contribution \eqref{omegaghostplanaire2} has an extra(2-d) IR singularity coming from the factor $\Gamma({{D}\over{2}}-1)$. Now, the UV finite non planar contribution \eqref{omegaghostnonplan} has an IR singularity coming from $(M^2)^{{{D}\over{2}}-2}=(M^2)^{-1}$, while the factor $({\sqrt{ {\tilde{p}}^2M^2}})^{2-{{D}\over{2}}}{{K}}_{2-{{D}\over{2}}}({\sqrt{ {\tilde{p}}^2M^2}})$ is IR finite. Using $K_1(z)\sim{{1}\over{z}}+(az+bz^3+...)+{{z^2}\over{2}}\log(z)+...$ where $a,\ b\in\mathbb{R}$, one checks that only the first term combines with $(M^2)^{-1}$ to give an IR singularity. Therefore, the small $|p|$ behaviour of \eqref{omegaghostnonplan} is
\begin{equation}
\omega_g^{NP}(p)\sim-{{1}\over{2\pi}}\int_0^1dxx{{1}\over{x(1-x)}}+\ IR\ regular,\ p\sim0\label{limnonplanghost},
\end{equation}
This IR singular term is exactly cancelled by the planar contribution \eqref{omegaghostplanaire2} at $D=2$:
\begin{equation}
\omega_g^{P}(p)={{1}\over{2\pi}}\int_0^1dx \, x \, {{1}\over{x(1-x)}}\label{limplanghost}.
\end{equation}
Hence the small $p$ limit of $\omega_g(p)$ is finite:
\begin{equation}
\lim_{p\to0}\omega_g(p)=\lim_{p\to0}(\omega_g^{P}(p)+\omega_g^{NP}(p))=finite\label{limitghost2ptir}.
\end{equation}
Similarly, the planar and non planar 1-loop contributions to the polarization tensor $\omega_{\mu\nu}(p)$ are
\begin{equation}
\omega^P_{\mu\nu}(p)={{1}\over{\pi}}\int_0^1dx{{1}\over{p^2x(1-x)}}(p^2\delta_{\mu\nu}-p_\mu p_\nu)\label{polarplan},
\end{equation}
\begin{equation}
\omega^{NP}_{\mu\nu}(p)=-2\int {{d^Dk}\over{(2\pi)^D}}{{\cos(p\wedge k)}\over{k^2(k+p)^2}}((2-D)[k^2\delta_{\mu\nu}-2k_\mu k_\nu]+2p^2\delta_{\mu\nu}-\frac{(D+2)}{2}p_\mu p_\nu)\label{nppolarisation}
\end{equation}
Setting {\footnote{From \eqref{JMU}, the 1st two terms between brackets in \eqref{nppolarisation} behave respectively as $\log|{\tilde{p}}|$ and ${{{\tilde{p}}_\mu{\tilde{p}}_\nu }\over{{\tilde{p}}^2 }}$, signaling UV/IR mixing unless $D=2$.}} $D=2$ and using \eqref{JN}, we obtain
\begin{equation}
\omega^{NP}_{\mu\nu}(p)=-{{1}\over{\pi}}\int_0^1dx{{1}\over{p^2x(1-x)}}\big({\sqrt{M^2{\tilde{p}}^2}}{{K}}_1({\sqrt{M^2{\tilde{p}}^2}})\big)(p^2\delta_{\mu\nu}-p_\mu p_\nu)\label{polarnonplan}.
\end{equation}
As above, by using the asymptotics given in the appendix \ref{feynmanappendix}, it can be checked that the  extra IR singularity in \eqref{polarnonplan} exactly cancels the planar contribution $\omega_{\mu\nu}^P$ \eqref{polarplan}. Hence
\begin{equation}
\omega_{\mu\nu}(p)=\pi(p^2)(p^2\delta_{\mu\nu}-p_\mu p_\nu),\ \ \lim_{p\to0}\pi(p^2)=finite \label{limitgauge2ptir}
\end{equation}
From \eqref{limitghost2ptir} and \eqref{limitgauge2ptir}, one concludes that IR singularities of 2-dimensional origin occurring in the 1-loop planar parts of ghost and gauge 2-point functions are exactly compensated by their non planar counterparts. This cancellation extends to the 3- and 4-point functions as  can be seen by using the Slavnov-Taylor identities \cite{hgw-13}. On general grounds, one can expect that these 2-d IR singularities depend on the gauge choice. This can be easily exemplified by considering the case of 1-loop 2-point functions for a commutative 2-d pure Yang-Mills theory,  for instance either in the Landau gauge or in the temporal gauge,  whose behavior is the same as the planar part of the above 2-point functions: IR singularities appear in the first gauge while they are simply absent in the other gauge. Notice that the present situation is slightly different since there are  no such IR singularities for both gauge choice. It suggests that the absence of these 2-d IR singularities does not  depend on the gauge choice. The coupling to a fermion has been discussed in \cite{hgw-13} and gives rise, as expected,  to a dynamical  mass generation mechanism as in the Schwinger model. \par 

\section{Noncommutative gauge theory as a matrix model}\label{section3}
Many attempts to deal with the induced gauge theory on Moyal space have tried to interpret it as a Yang-Mills type theory. Indeed, the action was assumed   to depend on the gauge potential $A_\mu$  or in other words the action \eqref{classaction} is considered as a functional of the gauge potential : $S=S[A_\mu]$. Promoting this interpretation beyond the classical level is still unsolved. In this section, we will change the above viewpoint and use the covariant field ${\cal{A}}_\mu$  defined in Eq. \eqref{covcoordinate} as the fundamental variable entering the action. Doing this, we will therefore formulate the induced gauge theory \eqref{inducedgaugematrix} as a matrix model $S=S[{\cal{A}}_\mu]$ and examine if such a matrix model invariant under ${\cal{A}}_\mu\rightarrow{\cal{A}}_\mu^g=g^\dag\star{\cal{A}}_\mu\star g,\ \forall g\in{\cal{U}}(\mathbb{R}^2_\theta)$ can have a consistent interpretation beyond the classical order. To our knowledge this interpretation has not been  explored so far, despite the  formal similarity between the first term of the action \eqref{inducedgaugematrix}   and the (bosonic part of) the action for some type IIB matrix models, such as the IKKT matrix model considered in \cite{blasch-stein-10}. \par 
\subsection{A family of gauge matrix models}\label{matrixmodelsfamily}

We will focus on the 2 dimensional version of \eqref{inducedgaugematrix}. We set
\begin{equation}
{\cal{A}}={{{\cal{A}}_1+i{\cal{A}}_2}\over{\sqrt{2}}},\quad {{\cal{A}}}^\dag={{{\cal{A}}_1-i{\cal{A}}_2}\over{\sqrt{2}}}.
\end{equation}
Then, one obtains
\begin{equation}
S_\Omega[\mathcal{A}]=\int d^2x\big((1+\Omega^2){\cal{A}}\star{\cal{A}^\dag}\star{\cal{A}}\star{\cal{A}}^\dag+(3\Omega^2-1){\cal{A}}\star{\cal{A}}\star{\cal{A}}^\dag\star{\cal{A}}^\dag+
2\kappa{\cal{A}}\star
{\cal{A}}^\dag\big)\label{classaction}.
\end{equation}
This action shares  some similarities with the 6-vertex model\footnote{We thank H. Steinacker for this remark.} although, as we shall see in the following, the entire analysis relies on the choice of a vacuum around which we shall perform fluctuations. We will come back to this  issue in a while.

It will be convenient to use the matrix base whose properties relevant for the ensuing analysis are recalled in  appendix \ref{matrixbase}. For more details, see \cite{Gracia-Bondia:1987kw}.\par

The strategy used in this section is standard, akin to the machinery of background field method used e.g in \cite{blasch-stein-10}: we choose a particular vacuum (the background), expand the action around it, fix the background symmetry of the expanded action. The equation of motion stemming from \eqref{classaction} is 
\begin{equation}
(3\Omega^2-1)({\cal{A}}^\dag\star{\cal{A}}\star{\cal{A}}+{\cal{A}}\star{\cal{A}}\star{\cal{A}}^\dag)+2(1+\Omega^2){\cal{A}}\star{\cal{A}}^\dag\star{\cal{A}}+2\kappa{\cal{A}}=0\label{eqmotion}.
\end{equation}
From now on, any solution of \eqref{eqmotion} will be denoted by $Z(x)$. In the following, we will consider solutions which respect the symmetries of the classical  action (symmetric solutions from now on), which were derived in \cite{GWW2}. They have the generic form 
\begin{equation}
Z_\mu(x)=\Phi_1(x^2)x_\mu+\Phi_2(x^2){\tilde{x}}_\mu
\end{equation}
where $\Phi_1$ and $\Phi_2$ are suitable functions that have to satisfy Eq. \eqref{eqmotion}. Passing to the matrix base, one can write equivalently
\begin{equation}
Z(x)= \frac{Z_1+i Z_2}{\sqrt{2}}=\sum_{m,n\in\mathbb{N}}Z_{mn}f_{mn}(x)
\end{equation}
where $Z_{mn}$ can be expressed in terms of $\Phi_1$ and $\Phi_2$.  This yields \cite{GWW2}
\begin{equation}
Z_{mn}=-ia_{m}\delta_{m+1,n}, \forall m,n\in\mathbb{N}\label{vacuummatrix}
\end{equation}
where  the sequence of complex numbers $\{a_m, m\in\mathbb{N}\}$ satisfies
\begin{equation}
a_m\big[(3\Omega^2-1)(|a_{m+1}|^2+|a_{m-1}|^2)+2(1+\Omega^2)|a_m|^2+2\kappa  \big]=0,\label{recursive}
\end{equation}
in view of \eqref{eqmotion}.\par 
This implies, $\forall m\in\ N$, $a_{-1}=0$,
\begin{itemize}
\item{$i)$} $a_m=0$
\item{$ii)$} $(3\Omega^2-1)(|a_{m+1}|^2+|a_{m-1}|^2)+2(1+\Omega^2)|a_m|^2+2\kappa=0$
\end{itemize}
The first solution corresponds to a trivial vacuum, whereas the second one engenders a whole family of symmetric vacua which depend on the range of values of the parameters $\Omega$ and $\kappa$. \par 
Setting formally ${\cal{A}}=Z+\phi$, ${\cal{A}}^\dag=Z^\dag+\phi^\dag$ in \eqref{classaction}, where $\phi$ can be interpreted as a fluctuation around $Z$, we obtain from the expansion of the Lagrangian, up to an unessential constant (we drop from now on the Moyal product symbol $\star$)
\begin{equation}
S[\phi,\phi^\dag]=S_2+S_3+S_4\label{expandaction}
\end{equation}
with
\begin{eqnarray}
S_2&=&\int d^2x\big[\phi^\dag\big(2(1+\Omega^2)ZZ^\dag+(3\Omega^2-1)Z^\dag Z\big)\phi+\phi\big(2(1+\Omega^2)Z^\dag 
Z+(3\Omega^2-1)Z Z^\dag\big)\phi^\dag \nonumber \\
&+&\phi^\dag\big((3\Omega^2-1)Z Z\big)\phi^\dag+\phi\big((3\Omega^2-1)Z^\dag Z^\dag\big)\phi \nonumber \\
&+&(3\Omega^2-1)(Z\phi^\dag Z^\dag\phi+Z^\dag\phi^\dag Z\phi)
+(1+\Omega^2)(Z\phi^\dag Z\phi^\dag+Z^\dag\phi Z^\dag\phi)+2\kappa\phi^\dag\phi\big]\label{quadraction},
\end{eqnarray}
\begin{eqnarray}
S_3&=&\int d^2x\big[(1+\Omega^2)\big(Z\phi^\dag\phi\phi^\dag+ Z^\dag\phi\phi^\dag\phi+Z\phi^\dag\phi\phi^\dag+Z^\dag\phi\phi^\dag\phi\big)\nonumber \\
&+&(3\Omega^2-1)(Z\phi^\dag\phi^\dag\phi+ Z^\dag\phi^\dag\phi\phi^\dag+Z^\dag\phi\phi\phi^\dag+Z\phi\phi^\dag\phi^\dag\big] \label{cubicaction}
\end{eqnarray}
\begin{equation}
S_4=\int d^2x \big[(1+\Omega^2)\phi\phi^\dag\phi\phi^\dag +(3\Omega^2-1)\phi\phi^\dag\phi^\dag\phi\big]\label{quartaction}.
\end{equation}
Notice that, when expanding around the trivial vacuum, the action \eqn{expandaction} simplifies considerably
\be
S^{(Z=0)}[\phi,\phi^\dag]=\int d^2 x \, [2\kappa \phi^\dag \phi +(1+\Omega^2) \phi \phi^\dag  \phi \phi^\dag + (3\Omega^2-1) \phi \phi^\dag \phi^\dag \phi]
\ee
This is the action of  a local matrix model, the so-called    6-vertex model\cite{Ginzparg} (also see \cite{Kostov} where the model has been solved). In the following  we shall concentrate on the non-trivial symmetric vacua which give rise to non-local models.
 
The action $S[\phi,\phi^\dag]$ in Eq. \eqref{expandaction} is invariant under a background transformation, that can be expressed through a nilpotent BRST-like operation, $\delta_Z$, with structure equations given by (star Moyal product is understood):
\begin{equation}
\delta_Z\phi=-i[Z+\phi,C],\ \delta_Z\phi^\dag=-i[Z^\dag+\phi^\dag,C],\ \delta_ZZ=0,\ \delta_ZC=iCC\label{brsbackgrd}.
\end{equation}
The background gauge symmetry can be fixed by adding to \eqref{expandaction} the following gauge-fixing action
\begin{equation}
S_{GF}=\delta_Z\int d^2x\ {\bar{C}}{\cal{G}}(\phi)=\int d^2x\ ( b{\cal{G}}(\phi))-{\bar{C}}\delta_Z{\cal{G}}(\phi))\label{gaugefixingphi},
\end{equation}
where the structure equations \eqref{brsbackgrd} have been supplemented with $\delta_Z{\bar{C}}=b$, $\delta_Zb=0$ and ${\cal{G}}(\phi)$ is the gauge function. The grading and ghost number assignments are as in the section 2. \par 

We find convenient to choose ${\cal{G}}(\phi)=(\phi-\phi^\dag)$ so that
\begin{equation}
S_{GF}=\int d^2x\ ( b(\phi-\phi^\dag)+i{\bar{C}}[Z-Z^\dag+\phi-\phi^\dag,C])\label{gaugefixinginterm}.
\end{equation}
In order to simplify slightly the situation, we now integrate over the $b$ field. In the BRST language, this amounts to consider a ``on-shell'' situation (for which now the nilpotency of the $\delta_Z$ operation is fulfilled modulo the ghost equation of motion together with the $\delta_Z$-invariance of the gauge-fixed action{\footnote{For a general discussion, see \cite{brs90}.}}). Doing this generates the constraint $\phi=\phi^\dag$ into \eqref{gaugefixinginterm} and \eqref{expandaction}. Then, \eqref{gaugefixinginterm} becomes
\begin{equation}
S_{GF}=\int d^2x\ i{\bar{C}}[Z-Z^\dag,C]\label{gaugefixingfinal}
\end{equation}
so that the ghosts decouple as expected. The action \eqref{expandaction} simplifies to a functional of $\phi$ only
\begin{equation}
S[\phi]=S_2+S_3+S_4\label{actfinal}
\end{equation}
with
\begin{eqnarray}
S_2=\int d^2x\ \big[\phi((1+5\Omega^2)(ZZ^\dag+Z^\dag Z)+(3\Omega^2-1)(ZZ+Z^\dag Z^\dag) \phi) \nonumber \\
+2(3\Omega^2-1)Z\phi Z^\dag\phi+(1+\Omega^2)(Z\phi Z\phi+Z^\dag\phi Z^\dag\phi)+2\kappa\phi\phi\big]\label{s2final},
\end{eqnarray}
\begin{equation}
S_3=8\Omega^2\int d^2x\ (Z+Z^\dag)\phi\phi\phi;\ S_4=4\Omega^2\int d^2x\ \phi\phi\phi\phi\label{s3s4final}.
\end{equation}
Plugging the expansion of $\phi$ in the matrix base, $\phi(x)=\sum_{m,n}\phi_{mn}f_{mn}(x)$, into \eqref{actfinal} gives rise to a matrix model of the form
\begin{equation}
S[\phi]=\sum_{m,n,k,l\in\mathbb{N}} \phi_{mn}\phi_{kl}G_{mn;kl}+S_{int}\label{actmatrix}
\end{equation}
where the kinetic operator is given by
\begin{eqnarray}
G_{mn;kl}&=&(1+5\Omega^2)\sum_{t\in\mathbb{N}}\delta_{ml}(Z_{nt}Z^\dag_{tk}+Z^\dag_{nt}Z_{tk})+(3\Omega^2-1)
\sum_{t\in\mathbb{N}}\delta_{ml}(Z_{nt}Z_{tk}+Z^\dag_{nt}Z^\dag_{tk})    \nonumber\\
&+&2(3\Omega^2-1)Z^\dag_{nk}Z_{lm}+(1+\Omega^2)(Z_{nk}Z_{lm}+Z^\dag_{nk}Z^\dag_{lm})+2\kappa\delta_{ml}\delta_{nk}\label{kineticoperator},
\end{eqnarray}
with $Z_{mn}$ given by \eqref{vacuummatrix}, $Z^\dag_{mn}=ia^*_m\delta_{n+1,m}$ and the cubic and quartic interaction terms are
\begin{equation}
S_{int}=8\Omega^2\sum_{m,p,q,r\in\mathbb{N}}i\phi_{pq}\phi_{qr}\phi_{mp}(a^*_r\delta_{m+1,r}-a_r\delta_{r+1,m})+
4\Omega^2\sum_{m,n,k,r\in\mathbb{N}}\phi_{mn}\phi_{nk}\phi_{kr}\phi_{rm}\label{interactmatrix}.
\end{equation}
Notice that the interaction terms both vanish whenever $\Omega^2=0$, as it should be with our gauge choice,   akin to the commutative situation for 2-d QCD in the axial gauge. Note also the occurrence of a cubic interaction term. \par
The kinetic operator \eqref{kineticoperator} can be rewriten as
\begin{eqnarray}
G_{mn;kl} &=&(1+5\Omega^2)\delta_{ml}\delta_{nk}(a_na^*_{n+1}+a^*_na_{n-1})\nonumber\\
&-&(3\Omega^2-1)(\delta_{ml}\delta_{n+1,k-1}a_na_{n+1}+\delta_{ml}\delta_{n-1,k+1}a^*_na^*_{n-1}
-2\delta_{m,l+1}\delta_{k+1,n}a^*_na_l)\nonumber \\
&-&(1+\Omega^2)(\delta_{k,n+1}\delta_{m,l+1}a_na_l+\delta_{n,k+1}\delta_{l,m+1}a^*_na^*_l)+2\kappa\delta_{ml}\delta_{nk}\label{kineticoperator1}
\end{eqnarray}
where the $a_m$'s are constrained by \eqref{recursive}. It can be observed that it involves 2 types of terms. The terms proportional to $(3\Omega^2-1)$ are non vanishing when $m+n=l+k\pm2$ while the remaining terms do not vanish whenever $m+n=l+k$ which is similar to the so called conservation law for indices that occurs in the matrix formulation of the Grosse-Wulkenhaar model \cite{grWm}.\par

\subsection{\texorpdfstring{The propagator at the special value $\Omega^2={{1}\over{3}}$}{Special value}}\label{omega13}

The propagator denoted by $P_{mn;kl}$ is defined by
\begin{equation}
\sum_{k,l}G_{mn;kl}P_{lk;sr}=\delta_{mr}\delta_{ns},\ \sum_{k,l}P_{nm;lk}G_{kl;rs}=\delta_{mr}\delta_{ns}\label{definvers}.
\end{equation}
It depends on the background which corresponds in each case to some solution of \eqref{recursive}. \par 

To simplify the ensuing discussion, we now assume that the $a_m$'s are real{\footnote{therefore setting the arbitrary phases $\xi_m$ to zero, in the notation of \cite{GWW2}}}. We also assume $\kappa\ne0$. The computation can be easily done when the terms proportional to $(3\Omega^2-1)$ are absent, i.e when $\Omega^2={{1}\over{3}}$. The corresponding solution of \eqref{recursive}, determined in \cite{GWW2} is given by:
\begin{equation}
\Omega^2={{1}\over{3}},\ \kappa<0,\ a_m={{1}\over{2}}{\sqrt{ -3\kappa}},\ \forall m\in\mathbb{N}\label{sol1}.
\end{equation}
Then, \eqref{kineticoperator1} becomes
\begin{equation}
 G^{(1/3)}_{mn;kl}=(-\kappa)\big(2\delta_{ml}\delta_{nk} - \delta_{k,n+1}\delta_{m,l+1}-\delta_{n,k+1}\delta_{l,m+1}\big)\label{kinetic13},
\end{equation}
and satisfies
\begin{equation}
G^{(1/3)}_{mn;kl}\ne0\iff m+n=k+l\label{conservation}.
\end{equation}
This conservation law acting on the indices implies that Eq.  \eqref{kinetic13} depends only on 2 indices. Indeed, setting $n=\alpha-m$, $k=\alpha-l$, with $\alpha=m+n=k+l$ into \eqref{kinetic13} yields
\begin{equation}
G^{(1/3)}_{m,\alpha-m;\alpha-l,l}:=G^{\alpha}_{m,l}=\mu^2(2\delta_{ml}- \delta_{m,l+1}-\delta_{l,m+1}),\ \forall m,l\in\mathbb{N}\label{tridiagon}
\end{equation}
where $\mu^2:=-\kappa$. Notice that it does not depend on $\alpha$. Therefore, we set $G^{\alpha}_{m,l}=G_{ml}$ to simplify the notations. \par 

One observes that $G_{ml}$ is an infinite real symmetric tridiagonal matrix which can be related to a Jacobi operator. Therefore, the diagonalization of \eqref{tridiagon} can be achieved by using a suitable family of Jacobi orthogonal polynomials. Note that a similar situation arises within the scalar Grosse-Wulkenhaar model \cite{grWm} as well as in noncommutative scalar field theory on $\mathbb{R}^3_\lambda$ constructed in \cite{vit-wal-12}. It is useful to recall here some technical points that will clarify the computation. For more mathematical details, see e.g \cite{orthop-1}. \par 
\subsubsection{Jacobi operators\label{Jacobi}}
A Jacobi operator $J$ acting on the Hilbert space $\ell^2(\mathbb{N})$ with canonical orthonormal basis $\{e_k \}_{k\in\mathbb{N}}$ can be defined as
\begin{eqnarray}
(Je)_k&=&a_ke_{k+1}+b_ke_k+a_{k-1}e_{k-1},\ k\ge1;\nonumber\\ 
(Je)_0&=&a_0e_1+b_0e_0 \label{jacoboperator},
\end{eqnarray}
where $\{a_k \}_{k\in\mathbb{N}}$and $\{b_k \}_{k\in\mathbb{N}}$ are infinite sequences of real numbers, with $a_k\ge0$, $\forall k\in\mathbb{N}$. It therefore can be represented as an infinite real symmetric tridiagonal matrix. Denoting by ${\cal{D}}(\ell^2(\mathbb{N}))$ the dense subset of $\ell^2(\mathbb{N})$ involving all finite linear combinations of the $e_k$'s, one can verify that $\langle JX,Y \rangle_{\ell^2}=\langle X,JY \rangle_{\ell^2}$, $\forall X,Y\in{\cal{D}}(\ell^2(\mathbb{N}))$ where $\langle,\rangle_{\ell^2}$ is the usual scalar product on $\ell^2(\mathbb{N})$. Then, $J$ extends to a densely defined symmetric operator on $\ell^2(\mathbb{N})$. \par 

If in addition $J$ is a bounded operator on ${\cal{D}}(\ell^2(\mathbb{N}))$, it extends by continuity to a self-adjoint (bounded) operator on ${\cal{D}}(\ell^2(\mathbb{N}))$. This occurs whenever 
\begin{equation}
\sup_k(|a_k|)+\sup_k(|b_k|)<\infty
\end{equation}
which is clearly verified by the operator defined by $J_{ml}=-G_{ml}$ from \eqref{tridiagon}. Then, as a corollary of the spectral theorem, the so called Favard theorem \cite{akhiez:1965} (see also \cite{orthop-1}) guaranties the existence of a set of (real) polynomials $\{p_k(x) \}_{k\in\mathbb{N}}$ orthogonal with respect to a unique compactly supported measure $d\mu(x)$, namely 
\begin{equation}
\langle p_m,p_n\rangle:=\int_\mathbb{R} d\mu(x)p_m(x)p_n(x)=\delta_{mn}\label{innerprodpolyn},
\end{equation}
and verifying the following  3-term recurrence relation:
\begin{eqnarray}
xp_k(x)&=&a_kp_{k+1}(x)+b_kp_k(x)+a_{k-1}p_{k-1}(x),\ k\ge1\nonumber
\\
xp_0(x)&=&a_0p_1(x)+b_0p_0(x)\label{recurencepolygen}.
\end{eqnarray}
Finally, provided some assumptions are verified \cite{akhiez:1965}, which will be the case below, the above scalar product \eqref{innerprodpolyn} can be expressed as the limit of a particular discrete inner product, namely 
\begin{equation}
\langle p_m,p_n\rangle=\lim_{N\to\infty}(\sum_{j=1}^Nw_j^2p_m(t_j)p_n(t_j))
\end{equation}
where the $t_j$'s and $w_j$ are respectively called the nodes and the weights. For more details see e.g \cite{akhiez:1965}. \par 

In the following, we will apply this formalism to the diagonalization of the kinetic operator in Eq. \eqn{tridiagon} through the determination of the relevant polynomials and related measure of integration. 

\subsubsection{Diagonalization of the kinetic operator}
Denoting generically by $\lambda_k$, $k\in\mathbb{N}$ the eigenvalues of $G_{mn}$ \eqref{tridiagon}, we write it as
\begin{equation}
G_{ml}=\mu^2\sum_{p\in\mathbb{N}}{\cal{R}}_{mp}\lambda_p{\cal{R}}^\dag_{pl}\label{unitarytrans}
\end{equation}
with
\begin{equation}
\sum_{p\in\mathbb{N}}{\cal{R}}_{mp}{\cal{R}}^\dag_{pl}=\sum_{p\in\mathbb{N}}{\cal{R}}^\dag_{mp}{\cal{R}}_{pl}=\delta_{ml}\label{orthogpoly},
\end{equation}
where ${\cal{R}}^\dag_{mn}={\cal{R}}_{nm}$. Then, the combination of Eqs. \eqref{kinetic13}, \eqref{unitarytrans} and \eqref{orthogpoly} gives rise to the following 3-term recurrence relation
\begin{equation}
{\cal{R}}_{m+1,q}+{\cal{R}}_{m-1,q}-(2-\lambda_q){\cal{R}}_{mq} =0,\ \forall m,q\in\mathbb{N}\label{recurmatrix},
\end{equation}
completed with
\begin{equation}
{\cal{R}}_{-1,q}=0,\ {\cal{R}}_{0,q}=1,\ \forall q\in\mathbb{N}.
\end{equation}
For further convenience, we set 
\begin{equation}
\rho_q=-\lambda_q,\ {\cal{R}}_{m}(\rho_q):={\cal{R}}_{mq}=,\ \forall q\in\mathbb{N}. 
\end{equation}
Then \eqref{recurmatrix} translates into
\begin{equation}
{\cal{R}}_{m+1}(\rho_q)+{\cal{R}}_{m-1}(\rho_q)=(2+\rho_q){\cal{R}}_{m}(\rho_q),\ \forall m,q\in\mathbb{N}\label{recurmatrixbis}
\end{equation}
with ${\cal{R}}_{-1}(\rho_q)=0$ and ${\cal{R}}_{0}(\rho_q)=1$ which stems from the following 3-term recurrence equation {\footnote{One more intuitive way to obtain the recurrence equation is to compute the secular equation for the matrix $G_{ml}$ \eqref{tridiagon} when $m=0$, $m\le1$, $m\le2$,... Calling $D_N(x)$, $N=m+1$, the secular equation, one observes by induction that $D_{N+1}(x)+D_{N-1}(x)=2xD_N(x)$ and in particular $D_1=2-x$, $D_2(x)=(2-x)D_1(x)-1$, $D_3(x)=(2-x)D_2(x)-D_1$ with $D_0=1$. }}:
\begin{eqnarray}
{\cal{R}}_{m+1}(x)+{\cal{R}}_{m-1}(x)&=&(2+x){\cal{R}}_{m}(x),\ \forall m\ge1\nonumber\\
{\cal{R}}_1(x)&=&(2+x){\cal{R}}_0(x) \label{chebyshev-gene}
\end{eqnarray}
with ${\cal{R}}_0(x)=1$ (${\cal{R}}_{-1}(x)=0$), evaluated at $x=\rho_q$, $q\in\mathbb{N}$.\par 
Eq. \eqref{chebyshev-gene} is of the type given by \eqref{recurencepolygen}. Now, we restrict ourselves to $N\times N$ submatrices, therefore we impose a cut-off on the indices, namely $0\le m,l,...\le N-1$ and define $J^N_{ml}:=(-G^N_{ml})$ in order to make contact with the  notation in \ref{Jacobi}. Then  Eq. \eqref{chebyshev-gene} can be cast into the matrix form (matrix product understood)
\begin{equation}
J^N\cdot \left(\begin{array}{c}
\mathcal{R}_0(x)\\
\mathcal{R}_1(x)\\ 
\cdots \\
\cdots \\
\mathcal{R}_{N-1}(x)
  \end{array}  \right) + \left(\begin{array}{c}
 0\\
0\\
\cdots \\ 
0 \\
{\cal{R}}_N(x)
  \end{array}  \right)=x\left(\begin{array}{c}
\mathcal{R}_0(x)\\
\mathcal{R}_1(x)\\ 
\cdots \\
\cdots \\
\mathcal{R}_{N-1}(x)
  \end{array}  \right)\label{finitematrixrecur}
\end{equation}
Eq. \eqref{finitematrixrecur} readily implies that the eigenvalues of $J^N$ are exactly given by the roots of ${\cal{R}}_N(x)$. Finally, setting $2t=2+x$ in \eqref{chebyshev-gene} yields 
\begin{equation}
U_{m+1}(t)+U_{m-1}(t)=2tU_m(t),\ \forall m\in\mathbb{N^*},\  U_{-1}=0,\ U_0(x)=1
\end{equation}
This defines the recurrence equation for the Chebyschev polynomials of 2nd kind \cite{kks}:
\begin{equation}
U_m(t):=(m+1){\mbox{$_2$F$_1$}(-m,\ m+2;\ {{3}\over{2}};\ {{1-t}\over{2}})}, \ \forall m\in\mathbb{N}\label{defcheb},
\end{equation}
where $_2$F$_1$ denotes the hypergeometric function. Note that the $U_m(x)$'s are a particular family of Jacobi polynomials \cite{kks} $P_n^{\alpha,\beta}(x)$, $U_m(t)={{P_m^{{{1}\over{2}},{{1}\over{2}}}(t) }\over{P_m^{{{1}\over{2}},{{1}\over{2}}}(1) }}$.\par
Putting all together, we can write
\begin{equation}
{\cal{R}}_m(x)=f(x)U_m({{2+x}\over{2}}),\ \forall m\in\mathbb{N}\label{solutionrecur},
\end{equation}
where the overall function $f(x)$ will be determined in a while so that \eqref{orthogpoly} holds true.\par 

The eigenvalues of $J^N_{ml}$ and therefore $G^N_{ml}$ are now entirely determined by the roots of $U_N(t)$. These{\footnote{ Recall first that, for any $N\in\mathbb{N}$, a Chebyshev polynomial of 2nd kind $U_N(t)$ has $N$ different simple roots in $[-1,1]$. }} are given by $t_k^N=\cos({{(k+1)\pi}\over{N+1}})$, $k=0,2,...,N-1$. Then, the eigenvalues for the kinetic operator $G^N_{ml}$ are 
\begin{equation}
\mu^2\lambda_k^N=2\mu^2(1-\cos({{(k+1)\pi}\over{N+1}})), \ k\in\{0,2,...,N-1\}\label{spectrumkinetic},
\end{equation}
and 
satisfy for finite $N$
\begin{equation}
0<\mu^2\lambda_N^k <4\mu^2\label{finiteeigenspec}.
\end{equation}
Summarizing our results, we have obtained:
\begin{equation}
{\cal{R}}^N_{mq}=f(N,q)U_m(t^N_q)=f(N,q){{\sin[{{\pi(m+1)(q+1)}\over{N+1}}] }\over{\sin[{{\pi(q+1)}\over{ N+1}}] }},\ 0\le m,q\le N-1\label{matrixrmq}
\end{equation}
where we used $U_m(\cos\theta)={{ \sin((m+1)\theta)}\over{\sin\theta }}$. \par 

To determine the normalization function $f(N,q)$, we first obtain from Eq.  \eqref{orthogpoly}
\begin{equation}
\sum_{p=0}^{N-1}{\cal{R}}^N_{pm}{\cal{R}}^N_{pl}=f(m,N)f(l,N)\sum_{p=0}^{N-1}U_p(t^N_m)U_p(t^N_l)\label{checkorthog}.
\end{equation}
The sum in the RHS of \eqref{checkorthog} can be computed by adapting the Christoffel-Darboux formula \cite{akhiez:1965} to the present situation. Indeed, multiplying the recurrence equation \eqref{recurencepolygen} for $p_m(x)$ by $p_m(y)$ as well as the recurence equation \eqref{recurencepolygen} for $p_m(y)$ by $p_m(x)$ and summing both, we obtain for any $x,y\in[-1,1]$.
\begin{eqnarray}
(x-y)\sum_{k=0}^{N-1}p_k(x)p_k(y)&=&p_N(x)p_{N-1}(y)-p_{N-1}(x)p_N(y),\ x\ne y\label{christoff-darb1},\\
\sum_{k=0}^{N-1}(p_k(x))^2&=&p^\prime_N(x)p_{N-1}(x)-p^\prime_{N-1}(x)p_N(x)\label{christoff-darb2},
\end{eqnarray}
where $f^\prime(x)$ denotes the derivative wrt $x$. Applying these relations to the $U_m$'s, \eqref{christoff-darb1} implies that \eqref{checkorthog} automatically vanishes whenever $m\ne l$ in view of $U_N(t^N_q)=0$. When $m=l$, we compute
\begin{equation}
\sum_{p=0}^{N-1}(U_p(t^N_m))^2=U_N^\prime(t^N_m)U_{N-1}(t^N_m)={{N+1}\over{((t^N_m)^2-1)}}(T_{N+1}(t^N_m)U_{N-1}(t^N_m))\label{relationinterm22},
\end{equation}
where the second equality in \eqref{relationinterm22} stems from the relation
\begin{equation}
U^\prime_N(x)={{(N+1)T_{N+1}(x)-xU_N(x) }\over{x^2-1 }}\label{relat-cheb2-5},
\end{equation}
and $T_N(x)$ denotes the $N$-th order Chebyshev polynomial of first kind \cite{kks}. By further using $T_N(\cos\theta)=\cos(N\theta)$ in \eqref{relationinterm22}, we finally arrive at
\begin{equation}
\sum_{p=0}^{N-1}(U_p(t^N_m))^2=(-1)^m(N+1){{ \sin[{{N(m+1)\pi }\over{N+1 }}  ]}\over {\sin^3[{{(m+1)\pi }\over{N+1 }} ] }}\label{relat-normalis}.
\end{equation}
One easily verifies that the RHS of \eqref{relat-normalis} is actually positive for any value of $m\in\mathbb{N}$. \par 
We finally obtain
\begin{eqnarray}
{\cal{R}}^N_{pm}&=&f(N,m)U_p(t^N_m)\label{finalrml1},\\
f(N,m)&=&\bigg((-1)^m(N+1){{ \sin[{{N(m+1)\pi }\over{N+1 }}  ]}\over {\sin^3[{{(m+1)\pi }\over{N+1 }} ] }} \bigg)^{-{{1}\over{2}}},\ 0\le p,m\le N-1\label{finalrml}.
\end{eqnarray}
Once we have the polynomials which diagonalize the kinetic term we can invert for the propagator. 
Keeping in mind Eqs. \eqref{definvers} and \eqref{tridiagon}, we set $P_{mn}:=P_{m,\alpha-n;\alpha-l,l}$ where $\alpha=m+n=k+l$. It follows from the above that for fixed $N$ the inverse of $G^N_{mn}$ denoted by $P^N_{mn}$ can be written as
\begin{equation}
P^N_{mn}={{1}\over{2\mu^2}}\sum_{p=0}^{N-1}f(N,p)^2U_m(t^N_p){{1}\over{1-t^N_p }}U_n(t^N_p)\label{propagfinite}.
\end{equation}
Taking the limit $N\to\infty$, the comparison of the relation $\delta_{ml}=\sum_p{\cal{R}}^N_{mp}{\cal{R}}^N_{lp}$ where the ${\cal{R}}^N_{mn}$'s are given by \eqref{finalrml1}, \eqref{finalrml} to the orthogonality relation among the Chebyshev polynomials $U_n$
\begin{equation}
\int_{-1}^1d\mu(x)\ U_m(x)U_n(x)={{\pi}\over{2}}\delta_{mn},\ d\mu(x)=dx{\sqrt{1-x^2}}\label{orthocontinucheb}
\end{equation}
permits one to trade the factor $f(N,p)^2$ (see \eqref{finalrml}) in $P^N_{mn}$ \eqref{propagfinite} for the compactly supported integration measure $d\mu(x)$ \eqref{orthocontinucheb} in \eqref{propagfinite}. \par 

To conclude this paragraph, we obtain the following rather simple expression for the inverse of the kinetic operator \eqref{tridiagon}
\begin{eqnarray}
P_{mn;kl}&=&\delta_{m+n,k+l}P_{ml},\nonumber\\
P_{ml}&=&{{1}\over{\pi\mu^2}}\int_{-1}^1dx\ {\sqrt{{{1+x }\over{ 1-x}}}}U_m(x)U_l(x)\label{propag-fin1}.
\end{eqnarray}
It can be easily checked that \eqref{definvers} is verified by combining \eqref{propag-fin1} with \eqref{kinetic13} and using the orthogonality relation \eqref{orthocontinucheb}. Notice that the integral in \eqref{propag-fin1} is well-defined leading to finite $P_{ml}$ when $m$ and $l$ are finite.\par 

\subsubsection{{Computation of the 1-point function}
}
Finally, 
let us consider briefly the 1-point function generated by the cubic vertex. The computation is standard. Supplementing $S[\phi]$ \eqref{actmatrix} by a source term $\sum_{m,n}\phi_{mn}J_{nm}$, the perturbative expansion can be obtained from the generating functional of the connected correlation functions $W[J]$ where
\begin{eqnarray}
{\cal{Z}}[J]&=&\int\prod_{m,n}d\phi_{mn}e^{-S[\phi]-\sum_{m,n}\phi_{mn}J_{nm}}=e^{W[J]}\label{generatingZ}\\
W[J]&=&\ln{\cal{Z}}(0)+W_0[J]+\ln\big(1+e^{-W_0[J]} ( e^{ -S_{int}(\frac{\delta}{\delta J})}   -1)e^{W_0[J]} \big)\label{connectgreen}\\
W_0[J]&=&\frac{1}{2}\sum_{m,n,k,l}J_{mn}P_{mn;kl}J_{kl}\label{connectfree}
\end{eqnarray}
and $S_{int}[\phi]$ collects the interaction terms of the action, which can be read off from Eq. \eqn{interactmatrix}, evaluated at  $\Omega^2=\frac{1}{3}$. The propagator  $P_{mn;kl}$ is given by Eq. \eqref{propag-fin1} and the $J_{nm}$'s are sources. Expanding the logarithm as a formal series yields all the connected diagrams while the effective action $\Gamma[\phi]$ is  obtained from $W[J]$ by Legendre transform
\begin{equation}
\Gamma[\phi]=\sum_{m,n}\phi_{mn}J_{nm}-W[J],\ \phi_{mn}=\frac{\delta W[J]}{\delta J_{nm}}\vert_{J=0}\label{effectiveaction}.
\end{equation}
From the formal expansion of \eqref{connectgreen} we obtain
\begin{eqnarray}
W^{1}[J]&=& \sum v_{mn}\big((P_{lm;kl}+P_{kl;lm})(P_{nk;cd}J_{cd}+J_{ab}P_{ab;nk})\nonumber\\
&+&(P_{lm;nk}+P_{nk;lm})(P_{kl;cd}J_{cd}+J_{ab}P_{ab;kl})\nonumber\\
&+&(P_{kl;nk}+P_{nk;kl})(P_{lm;cd}J_{cd}+J_{ab}P_{ab;lm}) \big),\label{w1}
\end{eqnarray}
with $\sigma=i\frac{2}{3}{\sqrt{3\mu^2}}$ and
\begin{equation}
v_{mn}=\delta_{m+1,n}-\delta_{m,n+1}\label{vmncubic}.
\end{equation}
Combining \eqref{w1} with the relevant term in the expansion solving the 2nd equation in \eqref{effectiveaction} given by
\begin{equation}
J_{mn}=\sum_{k,l}G_{nm;kl}\phi_{kl}+...\label{phiexpansion}.
\end{equation}
where $G_{nm;kl}$ is given by \eqref{kinetic13}, we obtain
\begin{eqnarray}
\Gamma^1[\phi]:=\sum_{r,s}\Gamma^1_{rs}\phi_{rs}&=&\frac{\sigma}{2}\sum v_{mn}\bigg((P_{lm;kl}+P_{kl;lm})(P_{nk;cd}G_{dc;rs}+G_{ba;rs}P_{ab;nk})\nonumber\\
&+&(P_{lm;nk}+P_{nk;lm})(P_{kl;cd}G_{dc;rs}+G_{ba;rs}P_{ab;kl}) \nonumber\\
&+&(P_{kl;nk}+P_{nk;kl})(P_{lm;cd}G_{dc;rs}+G_{ba;rs}P_{ab;lm})\bigg)\phi_{rs}\label{gammalinear}.
\end{eqnarray}
Now combining \eqref{gammalinear} with \eqref{propag-fin1} and
\begin{equation}
(P_{lm;cd}G_{dc;rs}+G_{ba;rs}P_{ab;lm})\phi_{rs}=2\phi_{ml}
\end{equation}
we can write $\Gamma^1[\phi]$  into the form
\begin{eqnarray}
\Gamma^1[\phi]&=&\sigma\sum_{m,n,k,l} v_{mn}\big(\delta_{mk}(P_{ll}+P_{km})\phi_{kn} +\delta_{nl}(P_{kk}+P_{nl})\phi_{ml}\nonumber\\
&+&(\delta_{m+l,n+k}P_{lk}+\delta_{n+k,m+l}P_{nm})\phi_{lk} \big).
\end{eqnarray}
By using the Kronecker delta symbols, $\Gamma^1[\phi]$ can be cast into the form
\begin{eqnarray}
\Gamma^1[\phi]=\sigma\sum(2P_{ll}-P_{l,l+1}+P_{kk}+P_{k+1,k+1}-P_{k,k+1})(\phi_{k,k+1}-\phi_{k+1,k})\label{effectonepoint}
\end{eqnarray}

It is divergent, with typical divergence 
\begin{equation}
\Gamma^1_{k,k+1}\sim\sum_l(2P_{ll}-P_{l,l+1}).
\end{equation}
This signals likely that the vacuum \eqref{sol1} is not stable against quantum fluctuations. Indeed, from the structure of $\Gamma^1[\phi]$ as in Eq. \eqref{effectonepoint}, it seems difficult (if possible at all) to absorb the divergences by the natural set of counterterms depending on two arbitrary parameters that are generated from the action \eqref{classaction} (when $\Omega^2=\frac{1}{3}$) and further expanded around the vacuum, akin to the linear-sigma model.\par
\section{Discussion and conclusion}{\label{discussion}}

A lot of information can be extracted from the propagator whose fate, and consequently the fate of the corresponding matrix model, is completely determined by the chosen vacuum. The UV and IR region can be identified from the spectrum of the kinetic operator. First, by taking the limit $N\to\infty$ of the spectrum given in \eqref{spectrumkinetic}, one observes that 
\begin{equation}
\lim_{N\to\infty}\mu^2\lambda_{k=0}^N=0,\ \lim_{N\to\infty}\mu^2\lambda_{k=N-1}^N=4\mu^2. 
\end{equation}
Then, from an overall rescaling of the initial action by a factor $g^{-2}$ with mass dimension $[g]=1$ so that $[\mu^2]=2$ (and $[{\cal{A}}]=1)$, it is natural to define the UV region as the one corresponding to large indices while the IR domain corresponds naturally to low indices, say, $m=0,1$. 

Next, we observe that the operator on $\ell^2(\mathbb{N})$ defined by the matrix elements $P_{ml}$  leads to a propagator which does not decay at large separation $|m-l|$. This can be already realized by expressing the matrix model in the field variables diagonalizing the kinetic operator, i.e the propagation base in the physics language. From \eqref{spectrumkinetic} and taking the large $N$ limit, one readily finds that the eigenvalues of the propagator $\mu^{-2}\lambda_k^\infty$ increase as the index $k$ increases which can be interpreted as signaling a lack of UV decay for the propagator. This can be verified directly on \eqref{propag-fin1} by computing numerically $P_{ml}$. Unsuppressed correlations $P_{ml}$ may exist at large separation $|m-l|$. This is to be compared with the situation for the Grosse-Wulkenhaar model for which there is a suppression of the correlations at large separation which then mimics a kind a quasi-locality, and was one of the main ingredients leading to the renormalizability of the model. Such a behaviour indicates that the  matrix model  under consideration, at the value $\Omega=\frac{1}{3}$ is highly non local.

We note that the interpretation of the action for the matrix model as the spectral action related to a finite volume spectral triple is not obvious, if possible at all. Such a triple, as introduced in \cite{harmonic-gw}, is characterized by a self-adjoint Dirac operator $D$ with compact resolvent operator whereas the defining algebra of the triple is not unital. In the present case, one should have $D^2=G$ ($G$ being the kinetic operator) from the spectral action computation. But $G$ is bounded, so that $G+\bbbone$ is also bounded. But then, $(G+\bbbone)^{-1}$ is not compact. Hence, the resolvent operator is not compact.

The lesson to be drawn is that requiring the propagator to decay at large indices, i.e in the UV, singles out one specific family of vacua among those symmetric vacua classified in \cite{GWW2}.  Let us analyze this point in some detail. In \cite{GWW2} the symmetric vacua have been classified according to  
\begin{eqnarray}
0<\Omega^2<\frac{1}{3}&,&\ u^1_m=\alpha(r^m-r^{-m})-\frac{\kappa}{4\Omega^2}(1-r^{-m}),\ \alpha\ge0,\ r>1\label{vac11}\\
\frac{1}{3}<\Omega^2<1&,&\ u^2_m=-\frac{\kappa}{4\Omega^2}(1-r^{-m}),\ \kappa\le0,\ r\le-1\label{vac12}\\
\Omega^2=1&,&\ u^3_m=-\frac{\kappa}{4}(1-(-1)^{-m})\label{vac13},\ \kappa\le0
\end{eqnarray}
with
\begin{equation}
r=\frac{1+\Omega^2+\sqrt{8\Omega^2(1-\Omega^2)}}{1-3\Omega^2}\label{err},
\end{equation}
($a_m=\sqrt{u_m}$ and t the arbitrary phase set to zero  as in the previous section), besides the solution 
 \eqref{sol1} for $\Omega^2=\frac{1}{3}$ that we have  analyzed in detail, which  leads to a propagator with an unsuitable decay. A similar conclusion applies to the 
solutions \eqref{vac12} and \eqref{vac13}. Indeed, one observes that $\lim_{m\to\infty}|u^2_m|=\frac{|\kappa|}{4\Omega^2}$ and $\lim_{m\to\infty}|u^3_m|\le\frac{|\kappa|}{2}$. In view of \eqref{kineticoperator1}, this corresponds to a bounded self-adjoint operator. Thus, the real spectrum, Spec$(G)$, is a compact subset of $[-||G||,||G||]$. This should not give rise to a decaying propagator. This observation does not apply to \eqref{vac11} for which the kinetic operator is unbounded, in view of 
\begin{equation}
\sqrt{u^1_m}=a^1_m\sim r^{\frac{m}{2}},\ m\to\infty\label{u1asymp}.
\end{equation}
Then, the corresponding propagator is  likely to have a suitable decay behaviour. This altogether singles out the family of symmetric vacua defined by \eqref{vac11}. However,  Eq. \eqref{u1asymp} indicates that the vacuum defined by Eq. \eqref{vac11} does not belong to the Moyal algebra, i.e the multiplier algebra recalled in  subsection \ref{use}. This would exclude the vacuum solution Eq. \eqref{vac11} if one insists to preserve the present definition of $\mathbb{R}^2_\theta$. Thus, one should conclude that the interpretation of Moyal gauge theory as a matrix model is  problematic, at least when using symmetric vacua. However, enlarging $\mathbb{R}^2_\theta$ to a new set including the solution \eqref{vac11} provides a way to overcome this obstruction. We think that this is an interesting possibility that deserves further investigations.\par

We notice that there is a formal similarity between the classical action \eqn{classaction} and the action describing the 6-vertex model  \cite{Ginzparg}. However, expanding \eqn{classaction} around a non-zero vacuum gives rise to a different model (the 6-vertex model  corresponds to a zero vacuum). Once the vacuum (background) is chosen, here among the symmetric non-trivial vacua of the classical action, the kinetic part is entirely determined, up to gauge fixing, by the part of the expanded action which is quadratic in the fluctuations. The kinetic part therefore depends on the vacuum solution chosen, e.g. \eqn{vacuummatrix}.\par

We have studied a class of noncommutative gauge theories elaborated in \cite{GWW} on 2-d Moyal space from the viewpoint of matrix models. We have explored some related properties beyond the classical order. Expanding the action around symmetric vacua classified in \cite{GWW2} generates non local matrix models with polynomial interaction terms. For a particular symmetric vacuum, we have shown that the kinetic operator is a Jacobi operator. The computation of the propagator has been carried out. The resulting propagator can be expressed in terms of Chebyschev polynomials of second kind. We have shown that non vanishing correlations exist at large separations.  For such  particular vacuum a divergent 1-point function appears that seems difficult to absorb by the set of symmetric counterterms, signaling possibly a loss of symmetry at the quantum level. The quantum stability of the vacuum is briefly discussed.  

From general spectral considerations on the kinetic operators stemming from the other classes of vacua determined in \cite{GWW2}, we have singled out a particular class of symmetric vacua that should lead to fast decaying propagators at large separation corresponding to quasi-local matrix models. The latter class of vacua is an interesting possibility to obtain consistent (quantum) matrix models from the present scheme and deserves further investigations. In particular, one should determine whether or not these vacuum solutions can be reconciled reasonably with algebraic structures related to Moyal noncommutative geometry and  compute the propagator from a non Jacobi kinetic operator.

\vskip 2 true cm
\noindent
{\bf{Acknowledgments}}: We warmly thank H. Steinacker for enlightening correspondence on recent developments in matrix models. Discussions with D. Blaschke and H. Grosse at various stages of this work are gratefully acknowledged. One of us (JCW) thanks H. Grosse for having pointed out past works related to spectral properties of operators.\par

\appendix
\section{\texorpdfstring{Vertex functions at $\Omega=0$}{Appendix}}\label{feynmanappendix}
It is convenient to define $p\wedge k\equiv p_\mu\Theta_{\mu\nu}k_\nu$ and $\tilde{p}_\mu=\Theta_{\mu\nu}p_\nu$. We list below the vertex functions {\footnote{Momentum conservation is understood. All the momenta are incoming}} for the Yang-Mills theory $\mathbb{R}^2_\theta$ ($\Omega=0$) in the Landau gauge. The propagator for the $A_\mu$ is $G_{\mu\nu}(p)=\delta_{\mu\nu}/p^2$. The ghost propagator is $G_{cc}(p)=1/p^2$.
\begin{equation}
V^3_{\alpha\beta\gamma}(k_1,k_2,k_3)=-i2\sin({{k_1\wedge k_2}\over{2}})\big[(k_2-k_1)_\gamma\delta_{\alpha\beta}+(k_1-k_3)_\beta\delta_{\alpha\gamma}+(k_3-k_2)_\alpha\delta_{\beta\gamma}\big]\label{v3},
\end{equation}
\begin{equation}
V^4_{\alpha\beta\gamma\delta}(k_1,k_2,k_3,k_4)=-4\big[ (\delta_{\alpha\gamma}\delta_{\beta\delta}-\delta_{\alpha\delta}\delta_{\beta\gamma})\sin({{k_1\wedge k_2}\over{2}})\sin({{k_3\wedge k_4}\over{2}}) \nonumber
\end{equation}
\begin{equation}
+(\delta_{\alpha\beta}\delta_{\gamma\delta}-\delta_{\alpha\gamma}\delta_{\beta\delta})\sin({{k_1\wedge k_4}\over{2}})\sin({{k_2\wedge k_3}\over{2}}) +(\delta_{\alpha\delta}\delta_{\beta\gamma}-\delta_{\alpha\beta}\delta_{\gamma\delta})\sin({{k_3\wedge k_1}\over{2}})\sin({{k_2\wedge k_4}\over{2}})\big]\label{v4},
\end{equation}
\begin{equation}
V_{cc\mu}(k_1,k_2,k_3)=i2k_{1\mu}\sin({{k_2\wedge k_3}\over{2}})\label{vcc}.
\end{equation}
The IR behaviour of the correlation functions can be conveniently controled by using the integrals given in \cite{WAL1}:
\begin{equation}
J_N({\tilde{p}})\equiv\int {{d^Dk}\over{(2\pi)^D}}{{e^{ik{\tilde{p}}}}\over{(k^2+m^2)^N}}=a_{N,D}{\cal{M}}_{N-{{D}\over{2}}}(m|{\tilde{p}}|)
\label{JN},
\end{equation}
\begin{equation}
J_{N,\mu\nu}({\tilde{p}})\equiv\int {{d^Dk}\over{(2\pi)^D}}{{k_\mu k_\nu  e^{ik{\tilde{p}}}}\over{(k^2+m^2)^N}}=a_{N,D}\big(\delta_{\mu\nu}{\cal{M}}_{N-1-{{D}\over{2}}}(m|{\tilde{p}}|)-{\tilde{p}}_\mu{\tilde{p}}_\nu{\cal{M}}_{N-2-{{D}\over{2}}}(m|{\tilde{p}}|) \big) \label{JMU},
\end{equation}
where
\begin{equation}
a_{N,D}={{2^{-({{D}\over{2}}+N-1)}}\over{\Gamma(N)\pi^{{D}\over{2}}}};\quad {\cal{M}}_Q(m|{\tilde{p}}|)={{1}\over{(m^2)^Q}}(m|{\tilde{p}}|)^Q{{K}}_Q(m|{\tilde{p}}|)\label{quantities}
\end{equation}
in which ${{K}}_Q(z)$ is the modified second kind Bessel function of order $Q\in{\mathbb{Z}}$. Recall that one has 
\begin{equation}
{{K}}_{-Q}(z)={{K}}_Q(z);\ \lim_{z\to0}z^\nu{{K}}_\nu(z)=2^{\nu-1}\Gamma(\nu),\ \nu>0\label{relationbessel1}
\end{equation}
so that the following asymptotic expansion holds true:
\begin{equation}
{\cal{M}}_{-Q}(m|{\tilde{p}}|)\sim2^{Q-1}{{\Gamma(Q)}\over{{\tilde{p}}^{2Q}}},\quad Q>0\label{asympt1}.
\end{equation}
\section{The matrix base\label{matrixbase}}
The matrix base used in the section \ref{section3} is the family of Wigner transition eigenfunctions of the harmonic oscillator \cite{Gracia-Bondia:1987kw} $\{f_{mn}(x)\}_{m,n\in\mathbb{N}}\subset{\cal{S}}$ which can be expressed as
\begin{equation}
f_{mn}={{1}\over{(m!n!\theta^{m+n})^{ {{1}\over{2}}} }}\bar{z}^{\star m}\star f_{00}\star z^{\star n}\label{fmngeneral}
\end{equation}
with $f_{00}=2e^{-{{1}\over{\theta}}z\bar{z}}$ and $a^{\star n}:=a\star a\star a\star...\star a$ (n times). The following relations hold true
\begin{equation}
(f_{mn}\star f_{pq})(x)=\delta_{np}f_{mq}(x),\ \int d^2x f_{mn}(x)=2\pi\theta\delta_{mn},\ f_{mn}^\dag(x)=f_{nm}(x),\ \forall m,n,p,q\in\mathbb{N}\label{basicdeffmn}
\end{equation}
Any element of the $\mathbb{R}^2_\theta$ can be writen as $a=\sum_{m,n}a_{mn}f_{mn}$ so that the $\star$-product between 2 elements of $\mathbb{R}^2_\theta$ is mapped into a product of 2 (infinite) matrices, namely $(a\star b)(x)=\sum_{m,n}c_{mn}f_{mn}(x)$, $c_{mn}=\sum_pa_{mp}b_{pn}$. Further usefull formulas are
\begin{eqnarray}
z\star\bar{z}^{\star n}\star f_{00}=n\theta\bar{z}^{\star(n-1)}\star f_{00}, f_{00}\star z^{\star n}\star\bar{z}=n\theta f_{00}\star z^{\star(n-1)}, n\ge1\label{computarelations}
\end{eqnarray}
while both RHS are zero for $n=0$.\par

\end{document}